\documentclass[aps, prd,twocolumn,superscriptaddress,preprintnumbers,floatfix,nofootinbib,notitlepage,showkeys,showpacs]{revtex4-1}

\usepackage[utf8x]{inputenc}
\usepackage[english]{babel}

\usepackage{graphicx}
\usepackage{hyperref}
\usepackage{amsmath}
\usepackage{amssymb}

\newcommand{\be}{\begin{equation}} 
\newcommand{\ee}{\end{equation}}
\newcommand{\bea}{\begin{equation}\begin{aligned}} 
\newcommand{\eea}{\end{aligned}\end{equation}}

\usepackage{color}

\begin{document}

\preprint{KCL-PH-TH/2018-68}

\title{Phase transition and vacuum stability in the classically conformal B--L model}

\author{Carlo Marzo}
\email{carlo.marzo@kbfi.ee}
\affiliation{NICPB, R\"avala 10, 10143 Tallinn, Estonia}  
\author{Luca Marzola}
\email{luca.marzola@cern.ch}       
\affiliation{NICPB, R\"avala 10, 10143 Tallinn, Estonia}          
\author{Ville Vaskonen}
\email{ville.vaskonen@kcl.ac.uk}
\affiliation{Physics Department, King's College London, London WC2R 2LS, United Kingdom}  

\begin{abstract}
Within classically conformal models, the spontaneous breaking of scale invariance is usually associated to a strong first order phase transition that results in a gravitational wave background within the reach of future space-based interferometers. In this paper we study the case of the classically conformal gauged B--L model, analysing the impact of this minimal extension of the Standard Model on the dynamics of the electroweak symmetry breaking and derive its gravitational wave signature. Particular attention is paid to the problem of vacuum stability and to the role of the QCD phase transition, which we prove responsible for concluding the symmetry breaking transition in part of the considered parameter space. Finally, we calculate the gravitational wave signal emitted in the process, finding that a large part of the parameter space of the model can be probed by LISA.
\end{abstract}

\maketitle

%-------------------------------------------------------------------------------
\section{Introduction}
%-------------------------------------------------------------------------------

The recent detection of the first gravitational wave signal by the LIGO collaboration~\cite{Abbott:2016blz} has opened a new observational window into the Universe. An important point of these investigations concerns the dynamics of phase transitions that occurred during the cosmological evolution, which may play a central role in an array of topics spanning from the problem of the baryon asymmetry of the Universe~\cite{No:2011fi,Huang:2016odd,Chala:2016ykx,Katz:2016adq,Artymowski:2016tme,Vaskonen:2016yiu,Dorsch:2016nrg,Beniwal:2017eik,Bian:2017wfv,Huang:2018aja,Beniwal:2018hyi} to the quest for an ultraviolet completion of the Standard Model (SM)~\cite{Schwaller:2015tja,Kakizaki:2015wua,Jinno:2015doa,Huber:2015znp,Leitao:2015fmj,Jaeckel:2016jlh,Dev:2016feu,Jinno:2016knw,Huang:2016cjm,Hashino:2016xoj,Kubo:2016kpb,Balazs:2016tbi,Baldes:2017rcu,Tsumura:2017knk,Demidov:2017lzf,Chao:2017vrq,Croon:2018erz,Hashino:2018zsi,Vieu:2018zze,Hashino:2018wee,Miura:2018dsy,Mazumdar:2018dfl,Brdar:2018num}. To provide a concrete example, gravitational wave astronomy has the potential to pinpoint the dynamics of the phase transition behind the generation of the electroweak scale, setting a new important benchmark for particle physics models. In fact, whereas the SM supports a second order electroweak phase transition, many of its extensions instead predict a first order phenomenon. In this case the electroweak phase transition proceeds through the nucleation and consequent expansion of bubbles that contain the true symmetry-breaking vacuum. Collisions between the bubbles and the motion in the plasma after bubble collisions then produce gravitational wave signals that can be detected in the present stochastic background by next-generation experiments such as the satellite-based interferometer LISA~\cite{Baker:2007}. 

In regard of this, classically conformal -- or scale-invariant -- models~\cite{Espinosa:2007qk,Espinosa:2008kw,Konstandin:2011dr,Konstandin:2011ds,Servant:2014bla,Fuyuto:2015vna,Sannino:2015wka} are an example of framework which typically induces a sizeable gravitational signature~\cite{Jinno:2016knw,Marzola:2017jzl,Iso:2017uuu,Hashino:2018zsi}, as thermal corrections here inevitably result in a potential barrier that separates the vacuum states of the theory. Presently the interest in conformal models has been revived for their possible connections with other problems in contemporary physics, involving for instance the origin of dark matter, the mechanism of cosmic inflation, vacuum stability or baryogenesis~\cite{Foot:2010av,Ishiwata:2011aa,Okada:2012sg,Heikinheimo:2013fta,Hambye:2013sna,Farzinnia:2013pga,Khoze:2013uia,Gabrielli:2013hma,Allison:2014zya,Allison:2014hna,Kannike:2014mia,Heikinheimo:2014xza,Kannike:2015apa,Karam:2015jta,Kannike:2015kda,Wang:2015cda,Marzola:2015xbh,Kannike:2016jfs,Kannike:2016wuy,Karam:2016rsz,Marzola:2016xgb,Okada:2012sg,Oda:2017zul,Helmboldt:2016mpi,Brdar:2018vjq}. 

In this work we continue these analyses by considering the classically conformal B--L model introduced originally in Refs.~\cite{Iso:2009ss, Iso:2009nw,Oda:2017kwl} and further studied in Refs.~\cite{Iso:2012jn,Oda:2015gna,Das:2015nwk,Das:2016zue,Khoze:2013oga,Guo:2015lxa}\footnote{We refer the reader to Refs.~\cite{Lindner:2013awa,Kanemura:2014rpa,Escudero:2016tzx,Coriano:2015sea, Escudero:2018fwn,Chauhan:2018uuy} for analyses of non-conformal B--L models.}. Differently from previous studies~\cite{Jinno:2016knw}, we pay particular attention to the impact of the SM QCD phase transition finding it responsible for accomplishing the electroweak symmetry breaking in part of the considered parameter space. The proposed analysis goes beyond the results of Ref.~\cite{Iso:2017uuu} by providing a first estimate of the bubble size at percolation, which allows for a reliable estimate of the gravitational wave signature of the model at hand. In line with the general results of Ref.~\cite{Ellis:2018mja}, we also find that thermal inflation~\cite{Lyth:1995hj,Lyth:1995ka} is a feature of the framework. In fact,  the contribution of the potential energy difference between true and false vacuum states is large enough to dominate the Hubble parameter at temperatures below the critical one, and the inflationary regime may last until the onset of the QCD phase transition with non-trivial consequences on additional phenomenology~\cite{Hambye:2018qjv,Baldes:2018emh}.  

The structure of the paper is as follows: after introducing the model in Sec.~\ref{sec:model}, we briefly discuss in Sec.~\ref{sec:constraints} its phenomenology at collider experiments and its impact on cosmology. The effective potential, including the contributions of thermal corrections and QCD phase transition, is presented in Sec.~\ref{sec:potential}, whereas the relative analyses of perturbativity and vacuum stability are detailed in Sec.~\ref{sec:rge}. The electroweak phase transition is studied in Sec.~\ref{sec:transition}, and the resulting gravitational wave signature of the model is computed in Sec.~\ref{sec:gws}. Finally, in Sec.~\ref{sec:conclusions}, we gather our conclusions.

%-------------------------------------------------------------------------------
\section{The model} \label{sec:model}
%-------------------------------------------------------------------------------

The model we consider is based on the symmetry group ${\rm SU}(3)_{\rm c}\times{\rm SU}(2)_{\rm L}\times{\rm U}(1)_{\rm Y}\times{\rm U}(1)_{{\rm B}-{\rm L}}$, with quarks and leptons having a B -- L charge of $+1/3$ and $-1$, respectively. The particle content of the SM is extended to include right handed neutrinos (RHN) $\nu_{Ri}$, required by the cancellation of the ${\rm U}(1)_{{\rm B}-{\rm L}}$ anomaly, and a complex scalar $\phi = (\varphi+iG)/\sqrt{2}$ that only carries a $+2$ ${\rm U}(1)_{{\rm B}-{\rm L}}$ charge. Notice that the SM Higgs doublet $H = (G_+, (h+iG_0)/\sqrt{2})$ transforms as a singlet under ${\rm U}(1)_{{\rm B}-{\rm L}}$. 

The scalar sector of the model is characterised by the following tree-level potential,
\be
V = \lambda_H (H^\dagger H)^2 + \lambda_\phi (\phi^\dagger \phi)^2 - \lambda_p (H^\dagger H)(\phi^\dagger \phi) \,,
\ee
where we include the so-called `portal coupling' $\lambda_p$ between the Higgs doublet and the new scalar field. 

In this setup, radiative corrections induce non-trivial solutions of the scalar potential minimization equation and consequently result in a symmetry breaking pattern which, generally, can be approximated in two subsequent stages: i)  radiative corrections produce an effective minimum of the potential along the $\varphi$ direction. The field $\varphi$ consequently develops a non-vanishing vacuum expectation value (VEV), $\langle\varphi\rangle\equiv w>0$, and the B--L symmetry is spontaneously broken. 
ii) The symmetry breaking dynamics is transmitted to the Higgs sector via a \textit{positive} portal coupling, which results in a negative mass term for the Higgs doublet: $\mu_H^2 = - \lambda_p w^2/2$. As a result, the electroweak symmetry is also spontaneously broken, and we have $\langle h\rangle\equiv v>0$ in concomitance with $w>0$. In Sec.~\ref{sec:potential} we discuss how a QCD phase transition prior to the B--L breaking dynamics changes this symmetry breaking pattern. 

We anticipate that our work will focus on the case where $w\gg v$, as collider bounds disfavour the complementary setup. The proposed symmetry breaking pattern is natural in this limit and matching the observed Higgs boson mass $m_h = 126\,{\rm GeV}$ and the electroweak VEV $v=246\,{\rm GeV}$ forces the Higgs boson quartic coupling approximately to its SM value, $\lambda_H \simeq m_h^2/(2v^2)$. The portal coupling is instead related to the VEV of $\varphi$ by $\lambda_p \simeq m_h^2/w^2$.

The interactions of the RHNs are given by 
\be
\label{eq:yuk}
- \mathcal{L}_\nu = Y_D^{ij} \, \overline{\nu_{Ri}} \tilde{H}^\dagger L_{Lj} + \frac{1}{2} Y_M^{ij} \, \phi \, \overline{\nu^c_{Ri}} \nu_{Rj} + {\rm h.c.}\,,
\ee
and after the symmetry breaking result in Dirac neutrino masses, as well as  Majorana masses for the RHNs. This implements the seesaw mechanism~\cite{1977PhLB...67..421M,1980PhRvL..44..912M,GellMann:1980vs,Yanagida:1979as} in the model, which ascribes the smallness of the measured (active) neutrino masses to a suppression factor given by the ratio between the neutrino Dirac mass scale and the RHN Majorana mass scale. In order not to clash with the bounds from Big Bang Nucleosynthesis we will consider RHNs with masses above 200 MeV.  

The final ingredient of the model is a kinetic mixing term for the ${\rm U}(1)_{\rm Y}$ and ${\rm U}(1)_{{\rm B}-{\rm L}}$ gauge fields, which is generally produced by quantum corrections even if set to zero at a scale. After diagonalising the kinetic term, the ${\rm U}(1)_{\rm Y}\times{\rm U}(1)_{{\rm B}-{\rm L}}$ part of the gauge covariant derivative is given by
\be
D^\mu \supset i g_{\rm Y} q_Y B_{\rm Y}^\mu + i (\tilde g q_{\rm Y} + g_{{\rm B}-{\rm L}} q_{{\rm B}-{\rm L}}) B_{{\rm B}-{\rm L}}^\mu \,,
\ee
where $\tilde g$ parametrizes the extent of the kinetic mixing and $q_j$, $g_j$ and $B_j^\mu$ are the charges, the gauge coupling and the gauge fields, respectively. Given the charge assignment of the extra scalar field $\phi$, the spontaneous symmetry breaking of the ${\rm U}(1)_{{\rm B}-{\rm L}}$ symmetry will induce a mass for the corresponding gauge boson given after the diagonalization by $m_{Z'}\simeq 2 g_{{\rm B}-{\rm L}} w$ for $w\gg v$. Notice also that in this case the $Z$-$Z'$ mixing angle is negligible. 

%-------------------------------------------------------------------------------
\section{Phenomenological consequences} \label{sec:constraints}
%-------------------------------------------------------------------------------

The interactions contained in Eq.~\eqref{eq:yuk} link the present framework to the problem of the origin of the baryon asymmetry detected in our Universe. In fact, as RHNs acquire a Majorana mass through the symmetry breaking of ${\rm U}(1)_{{\rm B}-{\rm L}}$, it is possible to implement the leptogenesis mechanism for baryogenesis. 

In standard scenarios of thermal leptogenesis~\cite{Buchmuller:2004nz,Fukugita:1986hr}, RHNs with hierarchical Majorana masses $M_i \gtrsim 10^9$ GeV are thermally produced in the plasma after inflation. As the Universe expands,  an original lepton, or B -- L, asymmetry is generated via the CP-violating out of equilibrium decays of the RHNs, and consequenlty partially converted into a baryon asymmetry by the SM sphaleron processes. Remarkably, possible pre-existing B -- L asymmetries can be efficiently washed out owing to the interplay between flavour effects and the RHN mass hierarchy~\cite{Bertuzzo:2010et,DiBari:2013qja,DiBari:2014eya}. In terms of the present analysis, implementing a standard leptogenesis scenario would force the $Z'$ mass scale well above the reach of contemporary collider experiments, being this parameter sourced by the same VEV $w$ behind the RHNs mass scale. On general grounds, we also expect sizeable coupling in Eq.~\eqref{eq:yuk} that could drive the running of the scalar sector parameters.    

As an alternative, it is possible to consider a scenario where RHNs with masses comparable to, or below, the electroweak scale, can produce the required baryon asymmetry via CP-violating flavour oscillations~\cite{Akhmedov:1998qx}. More in detail, the complex non-diagonal Majorana Yukawa matrices in Eq.~\eqref{eq:yuk} induce CP-violating $\nu_{Ri} \leftrightarrow \nu_{Rj}$ transitions, which conserve the overall lepton number but violate the lepton number of individual flavours. In this way, provided that at least one species of RHNs remains out of equilibrium while the SM sphalerons are active, the lepton asymmetry transmitted to the SM by the remaining RHN species will be reprocessed by the same sphaleron processes. Within the context of the conformal B -- L model, Ref.~\cite{Khoze:2013oga} adopted this mechanism to explain the observed  baryon asymmetry of the Universe. 

Finally, a third possibility to generate the required baryon asymmetry relies instead on $L$-violating Higgs decay, viable for right-handed neutrinos with masses below the Higgs boson mass~\cite{Hambye:2016sby, Hambye:2018qjv}. In this case the asymmetry is dominantly produced immediately before the offset of sphaleron processes. 

In the present work we implicitly assume either of the last two scenarios in order to simplify our computations. In fact, given the allowed RHN mass range, we can estimate the typical size of the neutrino Dirac Yukawa couplings via the seesaw formula,
\begin{equation}
    y_D\simeq \sqrt{\frac{m_\nu M_\nu}{v^2}}\lesssim10^{-5}
\end{equation}
where $M_\nu$ and $m_\nu$ are the mass scale of RH and active neutrinos, respectively. As a consequence, we can safely neglect the role of these parameters in determining the radiative corrections to the scalar sector. The same holds for the remaining couplings in Eq.~\eqref{eq:yuk}, provided that RHN do not exceed the electroweak scale:
\begin{equation}
	\label{eq:ymbound}
    y_M = 4 g_{{\rm B}-{\rm L}} \frac{M_\nu}{m_{Z'}}\lesssim 0.1 \,g_{{\rm B}-{\rm L}}\,. 
\end{equation}
For the above estimate we adopted a conservative bound, $m_{Z'}\gtrsim 4 $ TeV, suggested by the current $Z'$ searches for $g_{{\rm B}-{\rm L}} \gtrsim 10^{-1}$~\cite{Das:2015nwk,Das:2016zue, Escudero:2018fwn}. The impact of future experiments on this bound can be gauged by considering the prospects for sequential $Z'$ models at collider with increased luminosity or center of mass energy. A machine with a luminosity of $3000$ $fb^{-1}$ and $\sqrt{s} = 13-15$ TeV \cite{Atlas:2019qfx} potentially results in a lower bound only a few TeV higher than the current limits. Differently, with a setup of $30$ $ab^{-1}$ and $\sqrt{s} = 100$ TeV \cite{Benedikt:2018csr} as considered for the future circular collider, the bound would increase by an order of magnitude and reach a $\sim 45$ TeV limit for the sequential $Z'$ case.

% {\color{blue} An increase in luminosity and energy would ask for an upgrade of such bound, as can be inferred by simulations, for a sequential $Z'$,  at integrated luminosity $3000$ $fb^{-1}$ and $\sqrt{s} = 13-15$ TeV \cite{Atlas:2019qfx}, and at $30$ $ab^{-1}$ and $\sqrt{s} = 100$ TeV \cite{Benedikt:2018csr}. While in the first case the lower bound would have to be raised of few TeV, the reach of FCC would substantially change the lower limit up to an order of magnitude ($\sim 45$ TeV for the sequential $Z'$).
% } 
%-------------------------------------------------------------------------------
\section{Effective potential} \label{sec:potential}
%-------------------------------------------------------------------------------

The full scalar potential selects a surface in the two-dimensional field space and both the fields develop a non-zero VEV at the minimum generated by radiative corrections. However, for $w \gg v$, the properties of the initial B -- L phase transition can be inferred by analysing the dependence on $\varphi$ only~\cite{Prokopec:2018tnq}. 

The one-loop finite temperature effective potential along the $\varphi$ direction is given by
\be \label{Veff1}
V_{\rm eff} = V_0 + V_T  \,,
\ee
where the one-loop RG-improved $T=0$ contribution is
\be
\label{eq:v0}
V_0 = \frac{\lambda_\phi(t)}{4} \varphi^4 \,.
\ee
The RG scale is chosen such that $t=\log(\varphi/\mu_0)$, where $\mu_0$ is a reference scale. By using the above expression, we find that the mass of the extra scalar approximately scales as $m_\varphi \simeq 0.4 g_{{\rm B}-{\rm L}} m_{Z'}$.

The finite temperature part is
\be
V_T = \frac{T^4}{2\pi^2} \sum_j k_j J_T(m_j(\varphi)^2 + \Pi_j(T)) \,,
\ee
where the sum runs over the B -- L gauge boson, RHNs, the scalar boson $\varphi$ and the Goldstone boson $G$, and $k_j$ indicates the intrinsic number of degrees of freedom ($k_{Z'}=3,\, k_i=1,\, k_\varphi=1=k_G,$). The field dependent masses are given by
\bea
& m_{Z'}(\varphi)^2 = 4 g_{{\rm B}-{\rm L}}(t)^2 \varphi^2\, \quad M_i(\varphi)^2 = Y_i(t)^2\varphi^2/2 \,, \\
& m_\varphi(\varphi)^2 = 3\lambda_\phi(t) \varphi^2\,, \quad m_G(\varphi)^2 = \lambda_\phi(t) \varphi^2 \,,
\eea
and the Debye masses by
\bea
& \Pi_{Z'}(T) = 4 g_{{\rm B}-{\rm L}}(t)^2 T^2\,, \\
& \Pi_\varphi(T) = \frac{T^2}{24}\left[ 24 g_{{\rm B}-{\rm L}}(t)^2 + 8 \lambda_\phi(t) + \sum_i Y_i(t)^2 \right] \,, \\
& \Pi_G(T) = \Pi_\varphi(T) \,.
\eea
Here, the thermal integral $J_T$ is defined as 
\be
J_T(x) = \int_0^\infty {\rm d}y\, y^2 \ln\left[1\mp e^{-\sqrt{x+y^2}}\right] \,,
\ee
with the negative sign for bosons and the positive for fermions. 

Notice that depending on the values of these parameters, the $U(1)_{{\rm B}-{\rm L}}$ breaking phase transition might not occur above the QCD phase transition temperature, $T_{\rm QCD} = \mathcal{O}(0.1\,{\rm GeV})$. In this case, the QCD phase transition induces an additional linear term for the Higgs field, $\sum_j y_j \langle \psi_j \bar\psi_j \rangle h/\sqrt{2}$, which consequently acquires a non-zero VEV, $v_{\rm QCD} \equiv \langle h\rangle = \mathcal{O}(0.1\,{\rm GeV})$~\cite{Iso:2017uuu}. In turn, the portal coupling then induces a negative mass term for $\varphi$, so that at $T<T_{\rm QCD}$ the effective potential along the $\varphi$ direction becomes 
\be
V_{\rm eff}^{T<T_{\rm QCD}} = - \frac{\lambda_p(t) v_{\rm QCD}^2}{4} \varphi^2 + V_{\rm eff}^{T>T_{\rm QCD}} \,,
\ee
where $V_{\rm eff}^{T>T_{\rm QCD}}$ is given by Eq.~\eqref{Veff1}. In our analysis we take $v_{\rm QCD} = T_{\rm QCD}=0.1\,{\rm GeV}$. It is then evident that the QCD phase transition dynamics effectively inverts the symmetry breaking pattern sketched in Sec.~\ref{sec:model} on the parts of the parameter space where the colour confinement precedes the B--L breaking. 

%-------------------------------------------------------------------------------
\section{Vacuum stability and perturbativity} \label{sec:rge}
%-------------------------------------------------------------------------------

The high-energy behaviour of a model can be inferred by studying the renormalization flow of its parameters. The requirement of desireable properties, such as stability and perturbativity, then generally result in further constraints on the low-energy parameter space of the framework under examination. 

Within the SM, for example, renormalization group methods indicate the allowed top quark and Higgs boson mass windows through the requirements of i) limited interaction couplings (perturbativity), and ii) the absence of scalar background configurations with energies below the EW one (vacuum stability)~\cite{Cabibbo:1979ay,Grzadkowski:1986zw}, which crucially depends on the value of the top quark mass. In regard of this, according to present measurements of this parameter, the SM vacuum is only metastable~\cite{Bezrukov:2012sa,Buttazzo:2013uya,Masina:2012tz,Alekhin:2012py,Bezrukov:2014ina}. 

Many extensions have been proposed in the attempt to overcome this puzzling feature of the SM. For instance the simplest SM$\times U(1)_{{\rm B}-{\rm L}}$ framework with explicit symmetry breaking has been investigated up to next-to-leading precision in Refs.~\cite{Basso:2010jm,Basso:2011na,Datta:2013mta,Chakrabortty:2013zja,Coriano:2014mpa,Coriano:2015sea,Accomando:2016sge,DiChiara:2014wha}. However, all these analyses confirm that the extra Yukawa terms introduced in this simple scenario generally worsen the overall high-energy behavior of the model, in spite of the stabilizing effect of scalar mixing and gauge couplings. An exception to these conclusion is provided by classical conformal models~\cite{Oda:2015gna,Das:2015nwk,Das:2016zue}, where the requirement of a radiative $U(1)_{{\rm B}-{\rm L}}$ breaking bounds the magnitude of the parameters in the Majorana neutrino mass matrix. 

In fact, as shown in Ref.~\cite{Iso:2009ss}, the presence of a radiatively generated minimum breaking the B--L symmetry can be inferred independently from the SM Higgs background. Along the $\varphi$ direction, the minimization of the effective potential reduces to a single scale problem, 
% \begin{align}
% \label{eq:v0c}
% V'(\varphi) =& \frac{d}{d \varphi}\,\left(\frac{\lambda_\phi(t)}{4} e^{4\int_0^{t} \gamma_\varphi} \,\varphi^4 \right) \\ 
% =& \frac{1}{4} \, e^{4\int_0^{t} \gamma_\varphi} \,\varphi^3 \left(\beta_{\lambda_\phi}(t) + 4 \lambda_\phi(t) \left(\gamma_\varphi(t) + 1\right) \right) = 0\, ,
% \end{align}
\be \label{eq:v0c}
	\frac{{\rm d} V_0}{{\rm d} \varphi} = \frac{1}{4} \,\varphi^3 \left(\beta_{\lambda_\phi}(t) + 4 \lambda_\phi(t)\right) = 0\,.
\ee
Notice that in writing Eq.~\eqref{eq:v0c} we assumed negligible values of the portal coupling $\lambda_p$, as required for $v\ll w$. As a consequence, scalar mixing cannot be invoked here to stabilize the SM vacuum. The analytical minimization of the effective potential, resulting in the boundary condition $\beta_{\lambda_\phi} + 4 \lambda_\phi = 0$ for $t = \log(w/\mu_0)$, ensures that the $h-$independent part of the effective potential reaches its minimum at $\langle\varphi\rangle = w$ and increases elsewhere. Therefore, considering also the smallness of $\lambda_p$, in the conformal \mbox{B -- L} model instabilities can only be generated along the Higgs direction. 

The sign of the second derivative of the effective potential in the $\varphi$ direction, as computed from the non-trivial solution of Eq.~\eqref{eq:v0c}, receives a positive contribution from the gauge sector and a negative one from the Majorana Yukawa couplings. Hence, for $w$ to be in correspondence of a minimum of the potential (or, equivalently, to prevent a tachyonic scalar mass), the RHN Majorana mass scale must satisfy $M_\nu < m_{Z'}/2^{1/4}$~\cite{Das:2015nwk} (enforced in the present case by Eq.~\eqref{eq:ymbound}), in line with similar bounds concerning the evolution of $\lambda_p$ and the following instability~\cite{Coriano:2014mpa,Coriano:2015sea}.

\begin{figure}[h]
	\centering
	\includegraphics[width=0.38\textwidth]{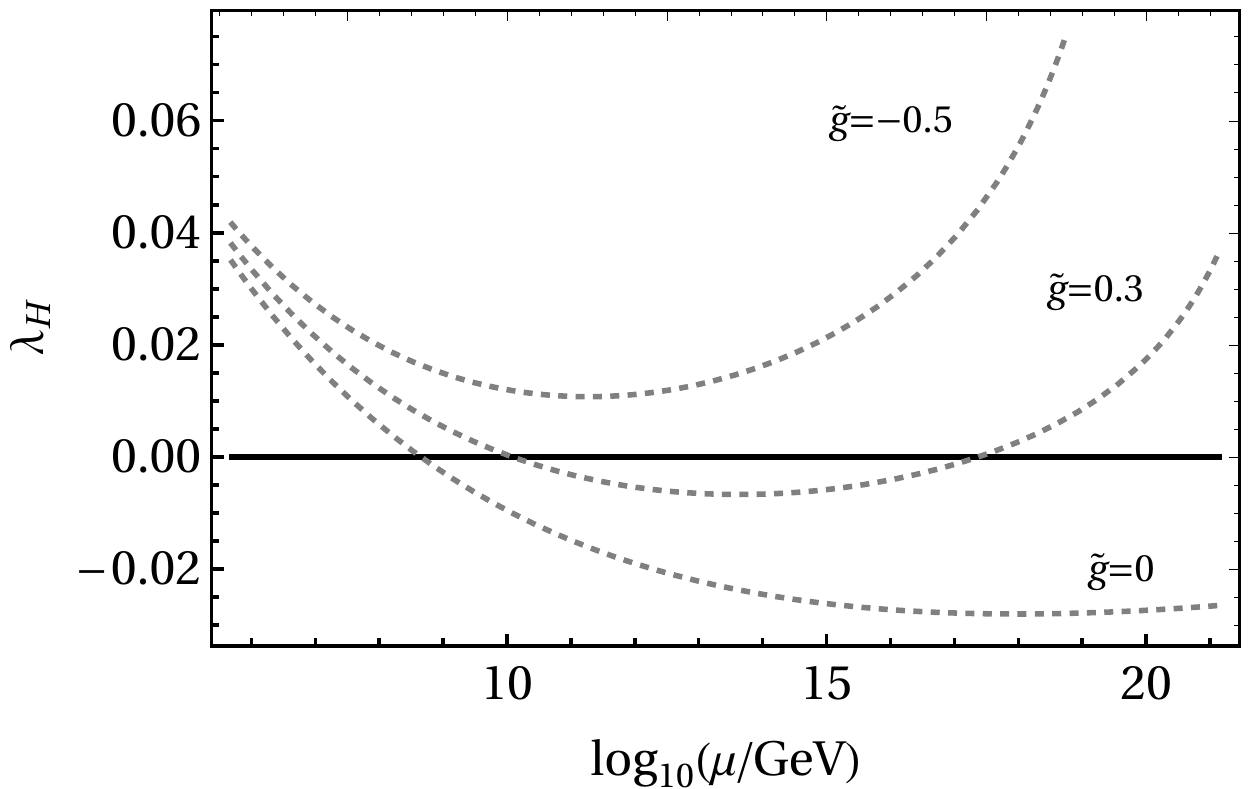}
	\caption{The evolution of $\lambda_H(t)$ as a function of the renormalization scale for $g_{{\rm B}-{\rm L}} = 0.1$, $m_{Z'} = 10^5$ GeV  and three different values of $\tilde{g}$. All couplings have been set to the indicated values at the $w$ scale.}
	\label{fig:lambdaH}
\end{figure}

Within the classical conformal case, the high-energy behavior of the model is therefore shaped by the extended gauge sector, which enters the RGE of the scalar sector through the gauge coupling $g_{{\rm B}-{\rm L}}$ and the mixing parameter $\tilde g$. The kinetic mixing, in particular, affects the evolution of $\lambda_H$ already at the one-loop level, allowing to solve the issue of the SM instability when the mixing is sizeable. This is explicitly shown in Fig.~\ref{fig:lambdaH}, where $\lambda_H$ is plotted as a function of the renormalization scale for three different values of $\tilde g$. The indicated values of the couplings have been set at the $w$ scale.

\begin{figure}
	\centering
	\includegraphics[width=0.37\textwidth]{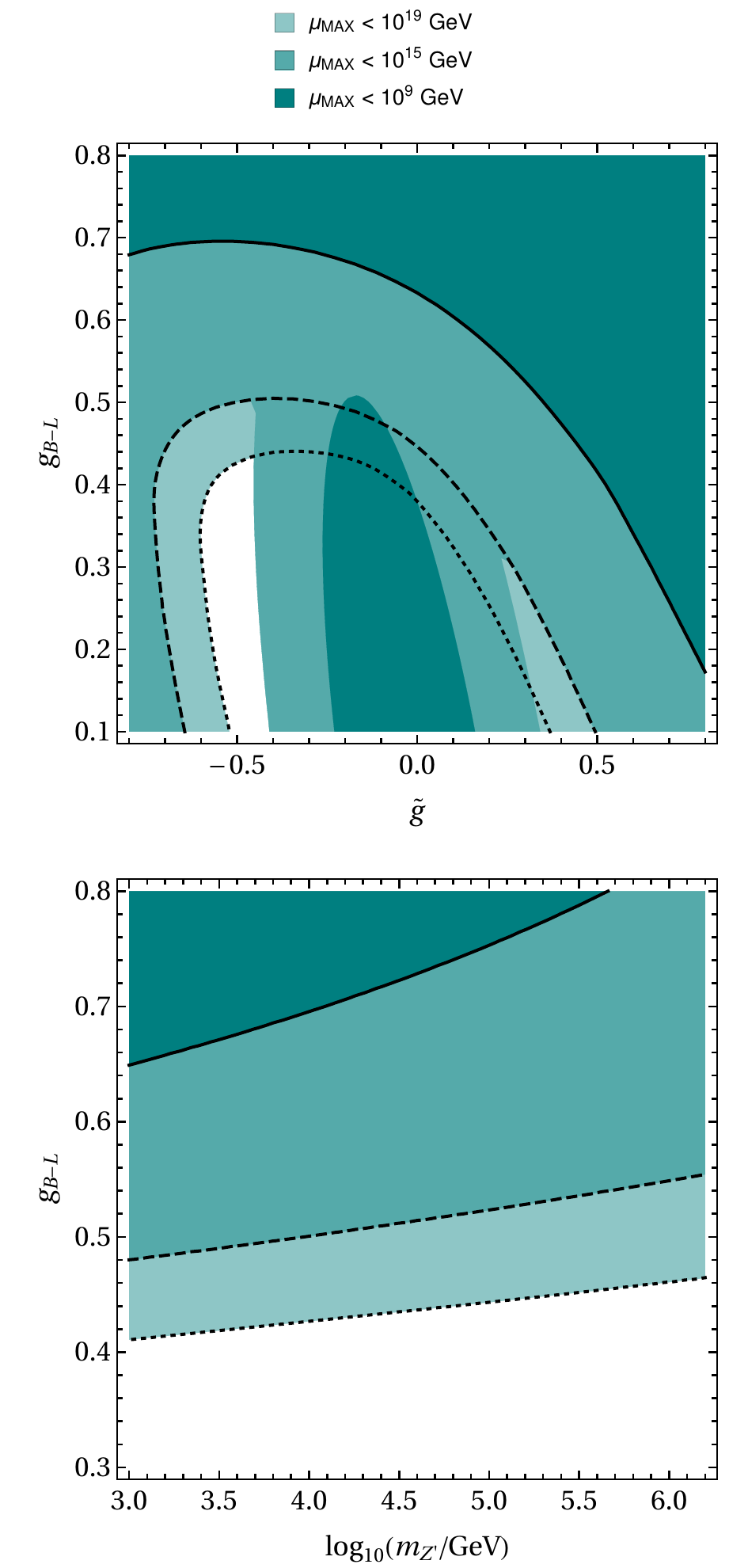}
	\caption{\emph{Top panel:} $M_{Z'} = 10$ TeV.  \emph{Bottom panel:} $\tilde{g} = -0.5$. In both the panels, coloured areas indicate the region of the parameter space where the stability of the symmetry-breaking vacuum is ensured up to the scale indicated in the legend. Below the dotted, dashed and solid lines we have the values of the parameters which allow the model to retain perturbativity of all the couplings (all couplings $< \sqrt{4\pi}$) beyond the Planck scale, at most up to the Planck scale and at most up to the GUT scale, respectively. Beyond the continuous black line, the model has a maximum perturbativity scale not exceeding the pure SM instability scale.}
	\label{fig:stabpert}
\end{figure}

The results of our analysis concerning the perturbativity and stability of the model are presented in Fig.~\ref{fig:stabpert} \footnote{Because of the hierarchy in the scales of the model, the RG equations have been solved by matching the SM evolution to the full model flow at the $Z'$ scale.}. The shaded areas in both the panels indicate the regions of the parameter space where the symmetry breaking vacuum of the model is stable up to three reference scales, which we chose as the instability scale of the pure SM ($10^9$ GeV), the grand unified theory (GUT) scale ($10 ^{15}$ GeV) and the Planck scale ($10^{19}$ GeV). Similarly, the perturbativity of the model (all couplings $<\sqrt{4\pi}$) can be retained well beyond the Planck scale in the region enclosed by the black dashed line. The dashed line and the solid line, instead, single out the parameter space where perturbativity is maintained until the Planck and GUT scales, respectively. On the outside of the solid contour, the perturbativity scale of the model does not exceed the scale of instability of the SM. As shown in the upper panel, the model is perturbative up to the Planck scale, and the electroweak vacuum is stabilised for $\tilde g \simeq -0.5$. We therefore adopt this value for the mixing parameter in the following analysis, anticipating that the phase transition dynamics do not significantly depend on this choice: $\tilde g$ does not directly affect the one-loop effective potential, entering only the running of $g_{{\rm B}-{\rm L}}$. To conclude the section, we remark that the sudden change in the stability of the potential for values of $\tilde g \gtrsim -0.4$ is due to the RG flow of $\lambda_h$ shown in Fig.~\ref{fig:lambdaH}.

%-------------------------------------------------------------------------------
\section{Phase transition} \label{sec:transition}
%-------------------------------------------------------------------------------

At very high temperatures, thermal corrections dominate the potential and  localize the fields at the origin, preventing the formation of new minima that would result in the spontaneous breaking of the symmetries of the model. This configuration is maintained until the temperature decreased enough to allow for the appearance of a second minimum in the potential, corresponding to a non-vanishing value of $\varphi$. We can therefore define the critical temperature $T_c$ as the temperature for which the new, symmetry-breaking, minimum is degenerate with the stationary point at the origin. We find numerically that 
\be
T_c \simeq 0.3 m_{\rm Z'} \gg T_{\rm QCD} \,.
\ee 
As the temperature further decreases, the symmetry-breaking minimum becomes a global minimum of the potential, but thermal corrections still result in a potential barrier that prevents the fields from leaving the origin. At temperatures $T \ll T_c$, the potential energy difference between the global minimum and the origin is then sizeable, whereas the height of the potential barrier progressively decreases. Quantum tunneling effects can then drive the fields to the global minimum of the potential, starting a first-order phase transition that proceeds via nucleation and consequent expansion of bubbles inside of which the symmetry is broken. 

The bubble nucleation rate per unit of time and volume can be estimated as~\cite{Linde:1981zj}
\be
\Gamma(T) \simeq T^4\left(\frac{S_3}{2\pi T}\right)^{\frac{3}{2}}\exp\left( -\frac{S_3}{T} \right) \,,	
\ee
where
\be \label{fullS3}
S_3 = 4\pi \int r^2 {\rm d}r \left[ \frac{1}{2}\left(\frac{{\rm d}\varphi}{{\rm d}r}\right)^2 + V_{\rm eff}(\varphi,T) \right]
\ee
is the action for an O(3)-symmetric bubble. The largest contribution into the above quantity arises from the classical path which minimizes $S_3$, corresponding to the solution of 
\be
\frac{{\rm d}^2 \varphi}{{\rm d} r^2} + \frac{2}{r} \frac{{\rm d} \varphi}{{\rm d} r} = \frac{{\rm d} V_{\rm eff}}{{\rm d} \varphi}
\ee
with boundary conditions ${\rm d}\varphi/{\rm d}r=0$ at $r=0$, and $\varphi\to0$ at $r\to\infty$. 

\begin{figure}[h]
\centering
\includegraphics[width=0.38\textwidth]{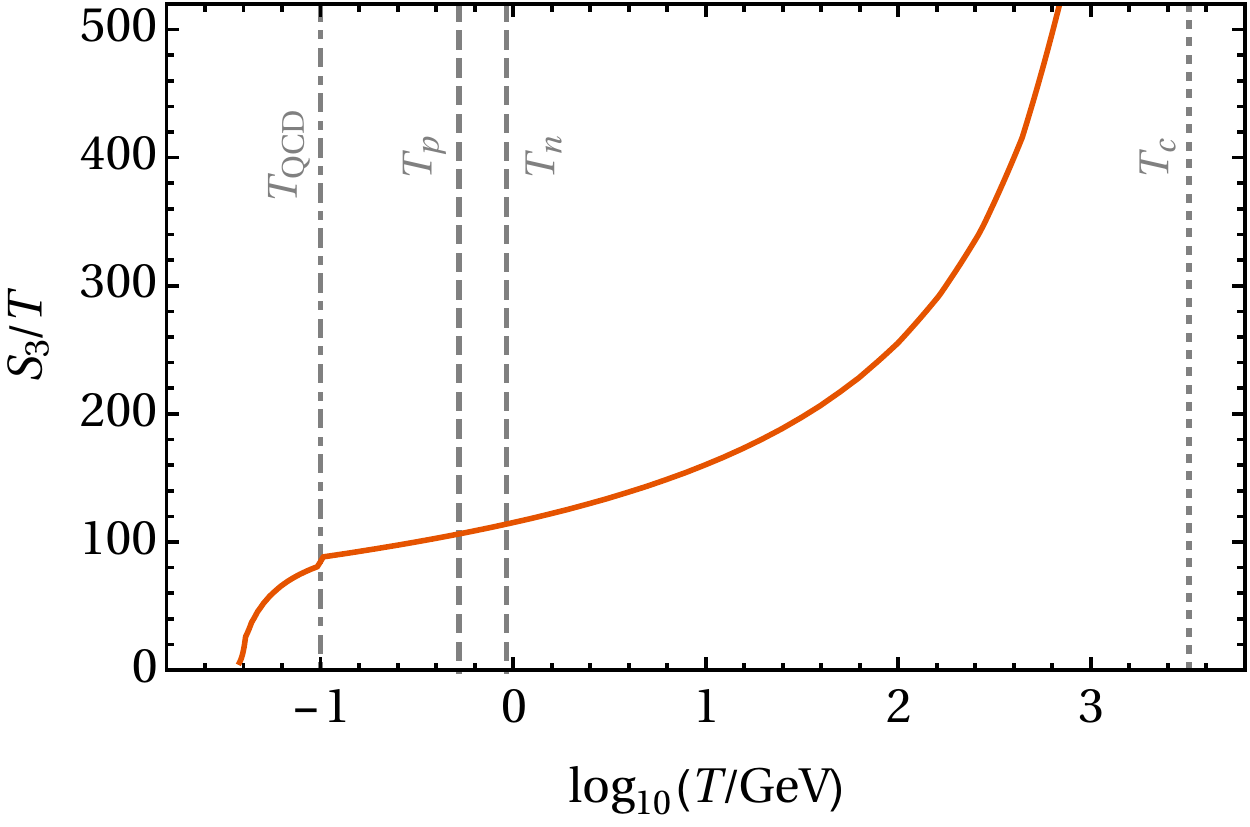}
\caption{The evolution of $S_3/T$ for a benchmark point with $m_{Z'} = 10\,{\rm TeV}$, $g_{{\rm B}-{\rm L}} = 0.26$, and $\tilde g(w) = -0.5$. The Majorana Yukawa couplings are assumed negligible.}
\label{fig:SoverT}
\end{figure}

The evolution of $S_3/T$ as a function of temperature is shown for a benchmark case in Fig.~\ref{fig:SoverT}. As temperature decreases, $S_3/T$ also decreases and eventually results in a sizeable bubble nucleation rate. However, below $T=T_{\rm QCD}$, the QCD phase transition changes the behaviour of $S_3/T$ inducing a negative mass term which cancels the thermal potential barrier, in a way that $S_3/T$ eventually vanishes.

We can then define the bubble nucleation temperature $T_n$ as the temperature at which the probability of producing at least one bubble per horizon volume in a unit of Hubble time approaches unity~\cite{Turner:1992tz},
\be
\frac{\Gamma(T_n)}{H(T_n)^4} \simeq 1 \,.
\ee
Notice that the Hubble rate $H$ includes the contribution due to the energy difference between the symmetry-conserving and symmetry-breaking vacua, $\Delta V(T=0)$. This vacuum energy dominates over the radiation contribution at
\be
T<T_v \simeq 0.3 T_c
\ee
as long as the phase transition is ongoing, and results in an epoch of thermal inflation.

Following Ref.~\cite{Ellis:2018mja}, we proceed by computing the volume fraction converted to the symmetry-broken phase at temperature $T$,
\be
I(T) = \frac{4\pi}{3} \int_T^{T_c} \frac{{\rm d} T' \,\Gamma(T')}{T'^4 H(T')} \left( \int_{T}^{T'} \frac{ {\rm d} \tilde T}{H(\tilde T)} \right)^3 \,.
\ee
The percolation temperature $T_p$ is defined as $I(T_p) = 0.34$, and here always satisfies the condition~\cite{Ellis:2018mja}
\begin{eqnarray}
    3+T {\rm d} I/{\rm d} T <0
\end{eqnarray}
enforcing that the physical volume of the patches still in the symmetric phase of the theory decrease. 

\begin{figure}[h]
\centering
\includegraphics[width=0.40\textwidth]{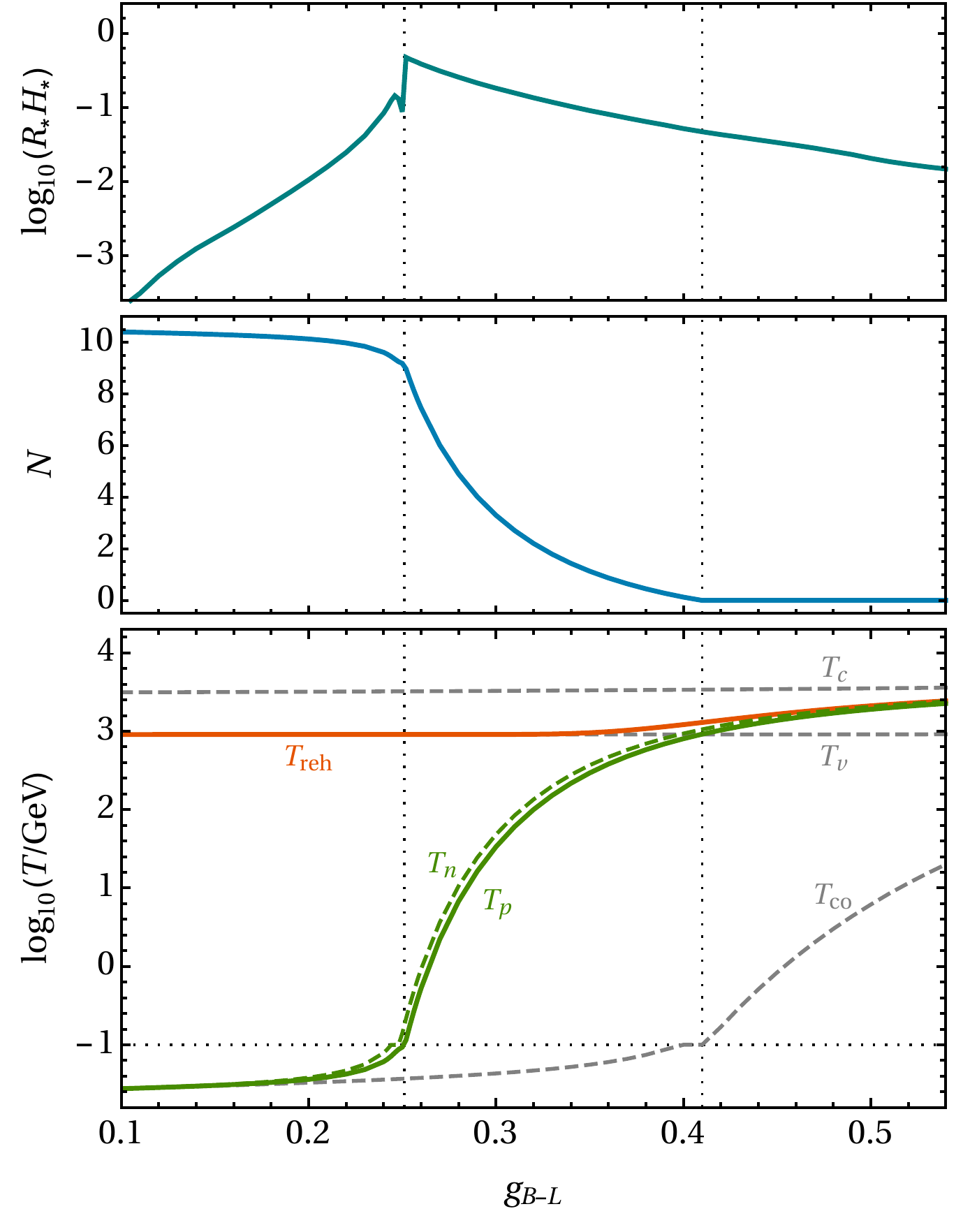}
\caption{The bottom panel shows the behaviour of the characteristic temperatures relevant for the phase transition (defined in the text) as a function of the B--L coupling. The middle and top panels show, instead, the number of e-folds of thermal inflation and the average bubble separation (at $T=T_p$ in Hubble lengths), respectively. On the left of the vertical dotted line at $g_{{\rm B}-{\rm L}} \simeq 0.42$, the phase transition completes during the vacuum dominated era, whereas on the left of the vertical dotted line at $g_{{\rm B}-{\rm L}} \simeq 0.25$ the phase transition dynamics conclude after the QCD phase transition. Here $m_{Z'} = 10\,{\rm TeV}$, $\tilde g(w) = -0.5$, and the Majorana Yukawa couplings are assumed negligible.}
\label{fig:TNRplot}
\end{figure}

The lines in the bottom panel of Fig.~\ref{fig:TNRplot} show the behaviour of the different temperatures defined above as a function of the B--L gauge coupling, assuming $m_{Z'} = 10\,{\rm TeV}$ and $\tilde g = -0.5$ and neglecting the Majorana Yukawa couplings. We see that percolation always takes place right after nucleation. The relative difference between the two corresponding temperatures is at its maximum immediately before the QCD phase transition starts to affect the dynamics. The temperature $T_{\rm co}$, represented by the lowest dashed line, signals the vanishing of the thermal potential barrier. For small values of $g_{{\rm B}-{\rm L}}$ this is due to the negative mass term induced by the QCD phase transition, whereas for larger $g_{{\rm B}-{\rm L}}$ this is caused by the running of $\lambda_\phi$.

The middle panel of Fig.~\ref{fig:TNRplot} shows instead that the number of $e$-folds of thermal inflation, $N \equiv \log(T_v/T_p)$, significantly increases for lower values of $g_{{\rm B}-{\rm L}}$. Consequently, only for $g_{{\rm B}-{\rm L}}\simeq 0.42$ the phase transition concludes during the radiation dominated epoch. We find that $N$ behaves as $N \simeq \log(m_{Z'}/{\rm GeV}) + C$ for $T_p < T_{\rm QCD}$, where $C$ is a constant that is mildly dependent on $g_{{\rm B}-{\rm L}}$ and $C(g_{{\rm B}-{\rm L}} = 0.1)\simeq1.0$. CMB studies then bound $m_{Z'}$ from above through the requirement that $N\lesssim60$, as required by the observed perturbation spectra.\footnote{The presence of a second period of inflation potentially reduces the required number of $e$-folds of primordial inflation indicated by the CMB measurements. For instance, the requirement $N\ll60$ yields values of the spectral index incompatible with observations within Starobinsky models~\cite{Starobinsky:1980te,Bezrukov:2008ej}, selecting alternative frameworks for the primordial inflation that predict a different dependence of this parameter on the number of $e$-folds, see e.g. Ref.~\cite{Dimopoulos:2016tzn}.} The bound however is not very efficient: for $g_{{\rm B}-{\rm L}} = 0.1$ we have $m_{Z'} \lesssim 10^{25}\,{\rm GeV}$.

After the phase transition concludes, the vacuum energy decays into radiation, consequently producing a plasma thermalised at a temperature $T_v < T_{\rm reh} < T_c$\footnote{Notice that there is no observational lower bound on the percolation temperature. Only after the phase transition, the reheating dynamics must bring the plasma to a temperature above the Big Bang nucleosynthesis one.} shown by the red line in the bottom panel of Fig.~\ref{fig:TNRplot}. We find that $T_{\rm reh}$ scales as $m_{Z'}$, and if $T_p \ll T_v$, the reheating temperature is $T_{\rm reh} \simeq T_v \simeq 0.09 m_{Z'}$. Given the parameter values indicated by collider experiments, the resulting temperature is high enough to restore the electroweak symmetry. This will be then broken in the same way as in the pure SM, as the temperature decreases below $T\simeq 140\,{\rm GeV}$ with the expansion of the Universe.

Our results concerning the phase transition dynamics are also presented in the first panel of Fig.~\ref{fig:PTplots}, as a function of $m_{Z'}$ and $g_{{\rm B}-{\rm L}}$. The color code indicates the percolation temperature, the dashed lines show the number of $e$-folds of thermal inflation, and the dot-dashed lines represent the reheating temperature. The thick black line highlights the contour $T_p = T_{\rm QCD}$, below which the transition happens only after the QCD one.

%-------------------------------------------------------------------------------
\section{Gravitational wave signal} \label{sec:gws}
%-------------------------------------------------------------------------------

We now discuss the gravitational wave signal emitted at the phase transition,  focusing on strongly supercooled dynamics, $T_p\ll T_c$, that occur during the vacuum energy dominance. We assume here that the bubble walls do not reach terminal velocity before they collide, as the energy density of the plasma is strongly depleted during the thermal inflation period. Bubble collisions then source the GW spectrum, which in the source frame is given in by~\cite{Cutting:2018tjt},\footnote{We remark that the numerical simulations in Ref.~\cite{Cutting:2018tjt} were performed for values of the bubble walls $\gamma$ factor $\gamma\sim\mathcal{O}(1)$, whereas the same parameters can be substantially larger in the present model. As recent studies indicate that large values of $\gamma$ may result in minor differences in the spectrum~\cite{Konstandin:2017sat,Jinno:2017fby}, we expect that the results of Ref.~\cite{Cutting:2018tjt} hold at least in order of magnitude.}
\be \label{gwspectrum}
\Omega_{\rm GW}(k) = (R_* H_* \Omega_v)^2 \frac{0.035 (k/{\tilde k})^3}{(1+1.99(k/{\tilde k})^{2.07})^{2.18}} \,,
\ee
as a function of the wave-number $k=2\pi\nu$. Here $\tilde k = 3.2/R_*$ corresponds to the peak frequency $\nu = \nu_{\rm env}$ of the spectrum, $H_* = H(T_p)$ is the Hubble rate,
\be
\Omega_v = \frac{8\pi \,\Delta V(T=0)}{3M_p^2 H_*^2}
\ee 
is the vacuum energy density parameter, and
\be
R_*^{-3} =  T_p \int_{T_p}^{T_c} \frac{{\rm d} T'}{T'^2} \frac{\Gamma(T')}{H(T')} e^{-I(T')}
\ee
is the average bubble separation~\cite{Ellis:2018mja} at the percolation temperature. Our computations indicate that $R_*$ scales roughly as $m_{Z'}^{-2}$, so the product $H_* R_*$ is basically independent of $m_{Z'}$. The dependence of $H_* R_*$ on the B$-$L gauge coupling is shown in the top panel of Fig.~\ref{fig:TNRplot}. Qualitatively, if the phase transition takes place above the QCD scale ($g_{{\rm B}-{\rm L}}\gtrsim 0.25$), we expect that decreasing values of $g_{{\rm B}-{\rm L}}$ lead to a progressively larger $H_* R_*$ because bubble nucleation rate decreases. On the contrary, if the transition occurs below the QCD scale ($g_{{\rm B}-{\rm L}}\lesssim 0.25$), decreasing values of $g_{{\rm B}-{\rm L}}$ enhance the bubble nucleation rate, which is driven here by the constant negative mass term induced by the QCD phase transition, leading to decreasing values of $H_* R_*$.

In order to predict the corresponding signal detectable at gravitational wave observatories, we let the gravitational waves emitted at the phase transition propagate until today. This amounts to a scaling of amplitude and frequency given by~\cite{Kamionkowski:1993fg}
\bea
&\frac{\Omega_{\rm GW}(T_0)}{\Omega_{\rm GW}(T_{\rm reh})} = 2.46\times 10^{-5} \left(\frac{100}{g_*}\right)^{1/3} \,, \\
&\frac{\nu(T_0)}{\nu(T_{\rm reh})} = 1.65\times 10^{-7} {\rm Hz} \,\frac{T_{\rm reh}}{H_*} \left(\frac{g_*}{100}\right)^{1/6} \,,
\eea
where $g_* = g(T_{\rm reh})$ is the effective number of relativistic degrees of freedom at the reheating temperature.

\begin{figure}
\centering
\includegraphics[width=0.38\textwidth]{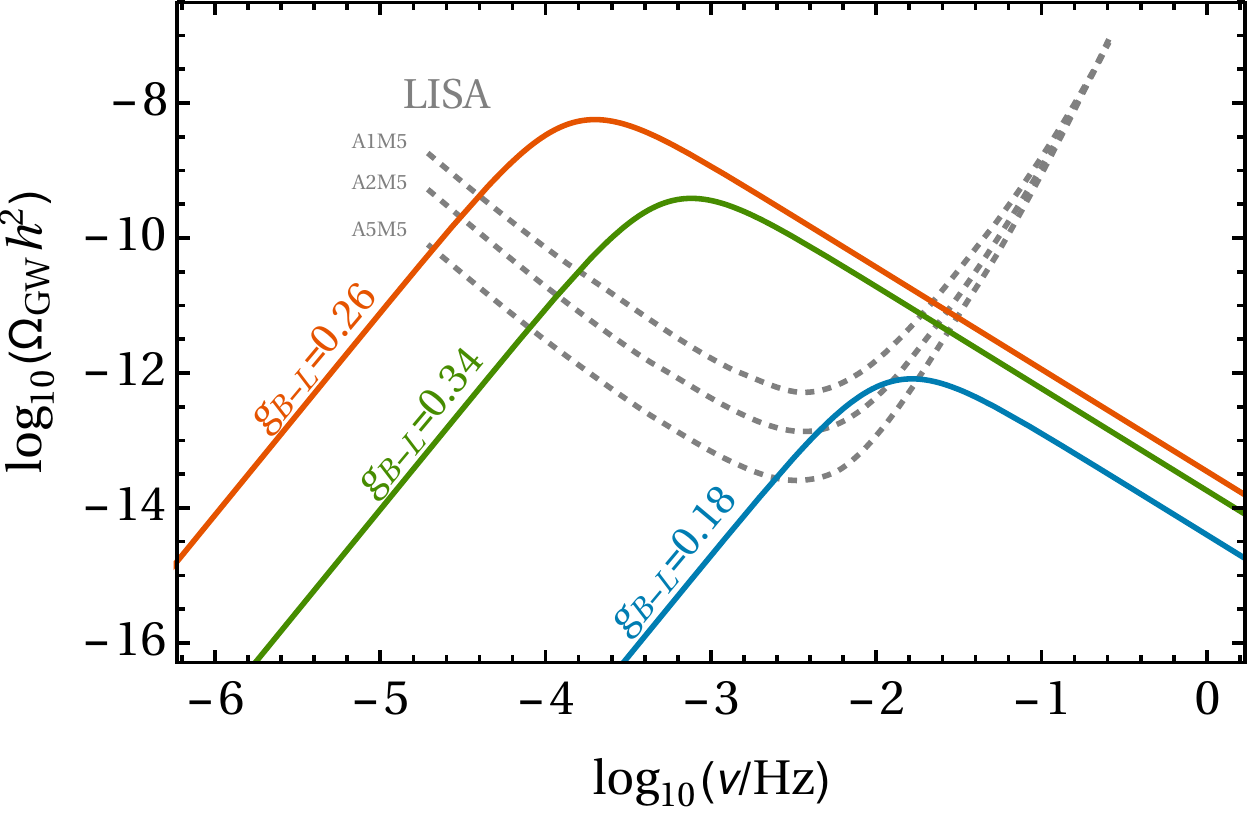}
\caption{Gravitational wave spectrum for three benchmark points. Here $m_{Z'} = 10\,{\rm TeV}$, $\tilde g(w) = -0.5$ and the Majorana Yukawa couplings are assumed to be negligible.}
\label{fig:gws}
\end{figure}

\begin{figure*}
\centering
\includegraphics[width=\textwidth]{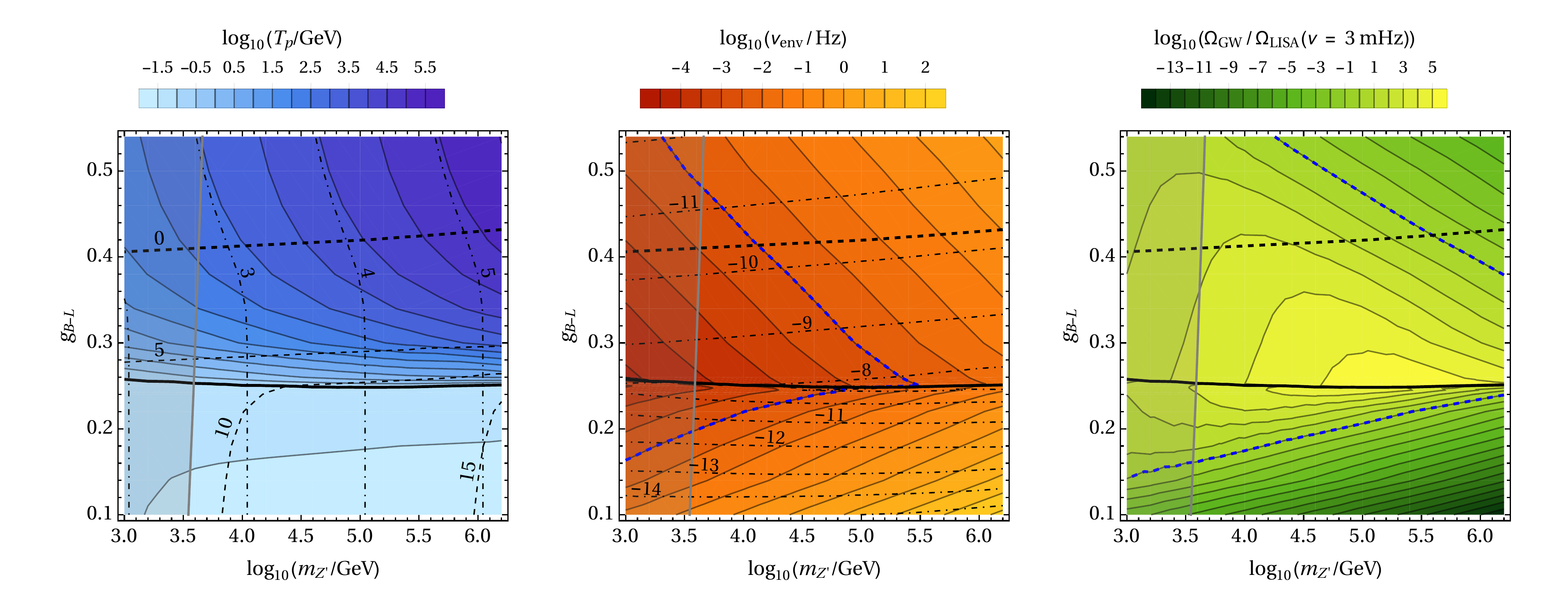}
\caption{Results shown in the $m_{Z'}$ -- $g_{{\rm B}-{\rm L}}$ plane for $\tilde g(w) = -0.5$. The Majorana Yukawa couplings are assumed to be negligible. In all panels the thick solid black line corresponds to $T_p=T_{\rm QCD}$ contour, and the thick dashed black line shows $N=0$. The region on the left of the gray line is excluded by the LHC searches. \emph{Left panel:} The color coding shows the percolation temperature, while the dashed lines show the duration in $e$-folds of the thermal inflation and the dot-dashed lines indicate the reheating temperature, $\log_{10}(T_{\rm reh}/{\rm GeV})$. \emph{Middle panel:} The color coding shows the peak frequency of the GW spectrum, and the dot-dashed lines indicate the maximal amplitude of the spectrum, $\log_{10}(\Omega_{\rm GW}(\nu=\nu_{\rm env}))$.  \emph{Right panel:} The color coding shows the amplitude of the GW spectrum relative to the LISA sensitivity at the frequency to which LISA is sensitive the most.}
\label{fig:PTplots}
\end{figure*}

The gravitational wave spectra generated for three benchmark points are shown in Fig.~\ref{fig:gws}. The signal is the strongest for $g_{{\rm B}-{\rm L}}=0.26$, as for smaller values of this parameter the phase transition takes place only after the QCD one. In fact, as shown in Fig.~\ref{fig:SoverT}, the slope of $S_3/T$ changes after the QCD phase transition has induced a negative mass term for $\varphi$, which effectively speeds up the process. This is manifest in the top panel of Fig.~\ref{fig:TNRplot}, showing the average bubble separation at $T_p$ as a function of $g_{{\rm B}-{\rm L}}$. 

Our final results are summarised in Fig.~\ref{fig:PTplots}. The left panel shows the percolation temperature, the number of $e$-folds of thermal inflation and the reheating temperature, as discussed in the previous section. In all panels the thick solid black line indicates where $T_p = T_{\rm QCD}$. Below this line, the phase transition happens after QCD has already induced a negative mass term for the $\varphi$ field. 

The middle and right panels of Fig.~\ref{fig:PTplots} characterise the GW emission consequent to the phase transition. The middle panel shows the peak frequency $\nu_{\rm env}$ and the amplitude of the spectrum at the corresponding frequency, $\Omega_{\rm GW}(\nu=\nu_{\rm env})$. We see that the strongest GW signal is obtained when the transition takes place immediately before the QCD one. The shape of the peak frequency contours follows from the $g_{{\rm B}-{\rm L}}$ dependence of $R_*$, shown in the top panel of Fig.~\ref{fig:TNRplot}, and the peak frequency increases as a function of $m_{Z'}$ as the spectrum is redshifted. The blue dashed line highlights the frequency $\nu=3\,{\rm mHz}$ to which LISA is sensitive the most. The mild dependence of $H_* R_*$ on $m_{Z'}$ is also evident from the behaviour of the $\Omega_{\rm GW}(\nu=\nu_{\rm enc})$ contours (dot-dashed lines).

Finally, our prospect for the detection at LISA of the GW spectrum emitted in the considered model is shown in the right panel of Fig.~\ref{fig:PTplots}. The color code indicates here the amplitude of the GW signal relative to the best sensitivity of the experiment. In the A5M5 setup, (see Fig.~\ref{fig:gws}), LISA would be able to probe the whole region between the blue dashed contours\footnote{Our analysis focuses on the LISA observational window, but  the future GW interferometers DECIGO and BBO will probe an even higher frequency range. As the peak frequency of the emitted GW signal increases with $m_{Z'}$, we expect that these experiments will exhaustively explore the parameter space of the model.}. 

Although GW foregrounds might not constitute an insurmountable obstacle for the detection of the predicted GW signal \cite{Adams:2013qma}, it is certainly true that most of the conformal extensions of the SM proposed in the literature result in comparable GW spectrums. Whether the non-observation of such gravitational signals could then question the realization of scale invariance in Nature, it is unlikely that their detection will alone reveal  the particle physics model behind the phase transition. Distinguishing between the proposed models therefore calls for complementary observations, provided in our case by possible detection of the $Z'$ peak at future colliders.

We remark that our prediction of the GW spectrum is \textit{not} to be trusted in the region above the black dashed line, where the phase transition concludes in the radiation dominated era. In fact, in this case Eq.~\eqref{gwspectrum} is not applicable as GWs originate from sound waves and turbulence in the plasma rather than bubble collisions. Notice however that entering such a region requires a substantial $g_{{\rm B}-{\rm L}}$ coupling, which according to the bottom panel of Fig.~\ref{fig:stabpert} sets the perturbativity scale of the model below the Planck scale.

%-------------------------------------------------------------------------------
\section{Conclusions} \label{sec:conclusions}
%-------------------------------------------------------------------------------
In this work we furthered the study of the conformal \mbox{B--L} extension of the Standard Model. After introducing the framework and briefly reviewing its general phenomenology, we focused on the phase transition dynamics that the scenario supports and on the high-energy properties of the theory.

With the RG-improved potential for the scalar sector of the theory at hand, we have identified a region in the parameter space of the model that ensures the stability of the potential and the perturbativity of its parameters up to scales well beyond the Planck one. In particular, we have found that the electroweak vacuum instability is here rescued by the effect of the gauge mixing, once the mixing parameter is set to  $\tilde g \simeq -0.5$.

Assuming this value in the following analysis (which is rather insensitive to  this parameter), we then studied the symmetry breaking pattern supported by the model, originated by the extra scalar field responsible for the radiative breaking of the B--L symmetry.  We find, in agreement with the earlier results of Refs.~\cite{Jinno:2016knw,Iso:2017uuu}, that thermal corrections prevent the transition to the emerging symmetry breaking minimum of the effective potential in a large part of the considered parameter space. As a consequence, we see the rise of an epoch of thermal inflation sourced by the potential energy difference between the false and true vacua of the theory. 

As originally pointed out in Ref.~\cite{Iso:2017uuu}, we have shown that  the inflationary regime concludes, at latest, soon after the onset of the QCD phase transition, when additional terms induced in the scalar potential weaken the thermal contribution and a first order electroweak phase transition takes place. The potential energy density of the false vacuum is then transferred to radiation in a second reheating process, which generally is strong enough to restore once again the electroweak symmetry. After the second reheating process concludes, depending on the achieved reheating temperature and the RHNs mass spectrum, the several leptogenesis mechanisms compatible with the framework give possibly rise to the observed baryon asymmetry of the Universe. The electroweak symmetry is instead finally broken after the temperature has dropped below the electroweak scale, in the same way as in the original Standard Model.

The main result of our investigation concerns the gravitational wave signature of the conformal \mbox{B--L} SM extension, emitted during the first order phase transition that follows the thermal inflation epoch. Being based on a direct estimate of the bubble size at percolation, our analysis improves on earlier studies~\cite{Jinno:2016knw,Iso:2017uuu} and delivers the expected gravitational wave signal over the whole parameter space of the model. We find that the amplitude of the spectrum is sizeable enough to fall within the reach of next-generation interferometers. In particular, LISA will probe most of the parameter space considered in the present analysis. 

\vspace{3mm}
\noindent
\textbf{Note added:} The assumption that the bubble walls do not reach terminal velocity has been closely scrutinized in Ref.~\cite{Ellis:2019oqb}, after the present paper was finalized. This new study has revealed that the assumption is justified as long as the B$-$L gauge coupling respects an upper bound of $g_{{\rm B}-{\rm L}}\simeq 0.25-0.3$ on the $Z'$ mass range considered in the present paper, with larger values of the parameters leading to a GW production sourced by the motion in the plasma rather than bubble collisions. In this regime, we find that the results shown in Fig.~\ref{fig:PTplots} still provide an order-of-magnitude estimate for the amplitude and the frequency of the signal.

%-------------------------------------------------------------------------------
\acknowledgments
We would like to thank Marek Lewicki for useful discussions. This work was supported by the United Kingdom STFC Grant ST/L000326/1 and by European Union through the ERDF CoE grant TK133.

\bibliography{citations}

\end{document}